\documentclass[eqsecnum,aps,prb,twocolumn,superscriptaddress,showpacs]{revtex4}% Physical Review B

\usepackage{graphicx}%
\usepackage{dcolumn}
\usepackage{amsmath} 

\makeatletter
\def\btt#1{\texttt{\@backslashchar#1}}%
\DeclareRobustCommand\bblash{\btt{\@backslashchar}}%
\makeatother

\begin{document}

\title[]{Fermi surface of the filled-skutterudite superconductor LaRu$_4$P$_{12}$: A clue to the origin of the metal-insulator transition in PrRu$_4$P$_{12}$}

\author{S.R.~Saha}
\email{sahasr@mcmaster.ca}

\altaffiliation{Present address: Department of Physics \& Astronomy, McMaster University,
Hamilton, Ontario L8S 4M1, Canada.}  

\author{H.~Sugawara}
\altaffiliation{Present address: Faculty of Integrated Arts and Sciences, Tokushima University, Tokushima 770-8502, Japan.}

\author{Y.~Aoki}
\author{H.~Sato}

\affiliation{Department of Physics, Tokyo Metropolitan University, Hachioji, Tokyo 192-0397, Japan}

\author{Y.~Inada}
\altaffiliation{Present address: Department of Science Education, Faculty of Education, Okayama University,
Okayama 700-8530, Japan.}

\author{H.~Shishido}
\author{R.~Settai}
\author{Y.~\={O}nuki}
\altaffiliation{Also at Advance Science Research Center, Japan Atomic Energy Research Institute, Tokai, Ibaraki 319-1195, Japan.}

\affiliation{Department of Physics, Osaka University, Toyonaka, Osaka 560-0043, Japan}

\author{H.~Harima}

\affiliation{Department of Physics, Kobe University, Nada, Kobe 657-8501, Japan}

\date{\today}

\begin{abstract}
We report the de Haas-van Alphen (dHvA) effect and magnetoresistance in 
the filled-skutterudite superconductor LaRu$_4$P$_{12}$, which is a reference
material of PrRu$_4$P$_{12}$ that exhibits
a metal-insulator (M-I) transition at $T_{\rm MI}\simeq60$~K. The observed dHvA branches for the main
Fermi surface (FS) are well explained by the band-structure calculation, using the 
full potential linearized augmented-plane-wave method with the local-density approximation, suggesting 
a nesting instability with ${\bf q}=(1,0,0)$ in the main multiply connected FS
as expected also in PrRu$_4$P$_{12}$. Observed
cyclotron effective masses of ($2.6-11.8)m_{\rm 0}$, which are roughly twice
the calculated masses, indicate the large mass enhancement even in the La-skutterudites.
Comparing the FS between LaRu$_4$P$_{12}$ and PrRu$_4$P$_{12}$, an essential role of
$c-f$ hybridization cooperating with the FS nesting in driving the the M-I transition
in PrRu$_4$P$_{12}$ has been clarified.
\end{abstract}

\pacs{75.20.Hr, 75.30.Mb, 71.18.+y, 71.27.+a}

\maketitle

%\section{Introduction}
%
The filled-skutterudite compounds $R$Tr$_4$Pn$_{12}$ ($R$ = rare earth,
Tr= Fe, Ru, Os; and Pn=pnictogen) have attracted much attention~\cite{sales} for 
exhibiting novel physical properties, i.e., metal-insulator (M-I) transition,~\cite{sekine1}
unconventional superconductivity,\cite{bauer,aoki1} semiconducting behavior,~\cite{torikak,sato1}
magnetic ordering,~\cite{torikak}
and their prospect in thermoelectric applications.~\cite{sales}
Particularly interesting are the Pr-based skutterudites in which quadrupolar interactions are believed to
play an important role for the anomalous properties.~\cite{bauer,sato1,aoki2,suga1,suga2} 
PrRu$_4$P$_{12}$ shows an exotic M-I transition
at $T_{\rm MI}\simeq60$~K,~\cite{sekine1} probably with a non-magnetic origin, 
as suggested by no anomaly in the magnetic susceptibility ($\chi$) 
and no essential effect of the magnetic field on the specific-heat
anomaly up to 12~T.~\cite{sekine1,sekine2} Although several studies have been
carried out,~\cite{sekine1,sekine2,lee1,lee2,hao1,saha1}
the mechanism of the M-I transition remains one of the most mysterious
puzzles in the skutterudites.

From the band-structure calculation,~\cite{suga3,harima1,harima2} Harima has suggested that the M-I transition
could be triggered by the nearly perfect nesting of the main Fermi surface (FS) with ${\bf q}$= (1,0,0); 
the shape of the main FS is nearly cubic and its volume is one-half of the body-centered cubic (bcc) Brillouin zone (BZ). Actually,
both the electron~\cite{lee1} and the x-ray~\cite{lee2,hao1}
diffraction experiments have revealed a doubling of the unit-cell volume
across $T_{\rm MI}$; it is suggested that the structure changes from bcc ($Im$\={3}) to simple cubic
($Pm$\={3}), caused by the different shift of Ru and P ions around two adjacent Pr ions, i.e,
between the corner and body-centered Pr ion sites in bcc structure. However, the predicted distortion of P-ion to
obtain an insulating band structure is much larger than that experimentally determined by the x-ray
diffraction.~\cite{harima2,curnoe,lee2} If the 4$f$ electrons in PrRu$_4$P$_{12}$ are
localized, as indicated by the Curie-Weiss behavior of $\chi$,~\cite{sekine1}
or only weakly hybridize with conduction electrons,
then the electronic structure is expected to be basically the same
as the reference material LaRu$_4$P$_{12}$,
which shows a superconducting transition at $T_{\rm C}\simeq7$~K,~\cite{shiro1} without a 4$f$ electron.
In that case, LaRu$_4$P$_{12}$ is also expected to have the same nesting instability,
however no M-I transition has been reported. 
Thus, it is  essentially important to clarify the FS of both compounds experimentally to understand the origin of the M-I transition in PrRu$_4$P$_{12}$. 
However, the high transition temperature of $T_{\rm MI}\simeq60$~K prevents the observation of dHvA
oscillations in the metallic state of PrRu$_4$P$_{12}$.
As an alternative way, the experimental determination of FS in LaRu$_4$P$_{12}$ should be useful to understand the origin of the M-I transition
in PrRu$_4$P$_{12}$, if it is carefully compared with the computed FSs of the two compounds based on the band-structure calculation.

We have succeeded in growing high-quality single crystals of LaRu$_4$P$_{12}$ and
in observing the dHvA effect, the preliminary result of which has been briefly reported 
in Ref. 20. In this paper, we report
the detailed experimental results on the dHvA effect and magnetoresistance along with
the band-structure calculation in LaRu$_4$P$_{12}$, followed by a discussion on the
possible origin of the M-I transition in PrRu$_4$P$_{12}$.

%\section{experiment}
Single crystals of LaRu$_4$P$_{12}$ were grown by the tin-flux method which
is basically the same
as described in Ref. \onlinecite{torikak}. The raw materials were
4N (99.99\% pure) -La, -Ru, 6N-P, and 5N-Sn.
The residual resistivity ratio (RRR) of the present sample is about 700, 
indicating the high sample quality.
The dHvA experiments were performed in a
top loading $^3$He cryostat down to 0.4 K with a 15~T superconducting (SC) magnet and a top loading dilution
refrigerator cooled down to $\sim30$~mK with a 17~T SC magnet. The dHvA signals were detected by 
means of the conventional field modulation method with a low frequency ($f\sim10$~Hz). The magnetoresistance 
was measured by the dc four-probe method using a top loading $^3$He cryostat equipped with a 16~T SC magnet. 

\begin{figure}[hbtp]
%h=here, t=top, b=bottom, p=separate figure page
\begin{center}\leavevmode
\includegraphics[width=6.5cm]{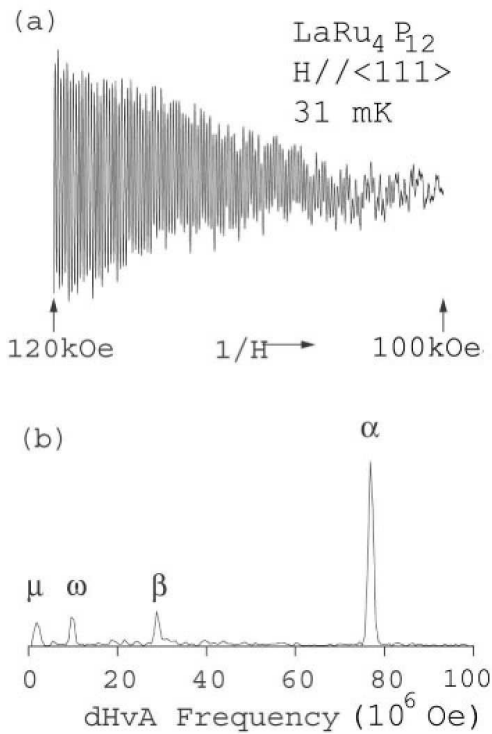}
\caption{(a) Typical dHvA oscillations and (b) its fast Fourier transformation (FFT) spectrum in LaRu$_4$P$_{12}$.}
\label{Osc_FFT}
\end{center}
\end{figure}

\begin{figure}[hbtp]
%h=here, t=top, b=bottom, p=separate figure page
\begin{center}\leavevmode
\includegraphics[width=8.5cm]{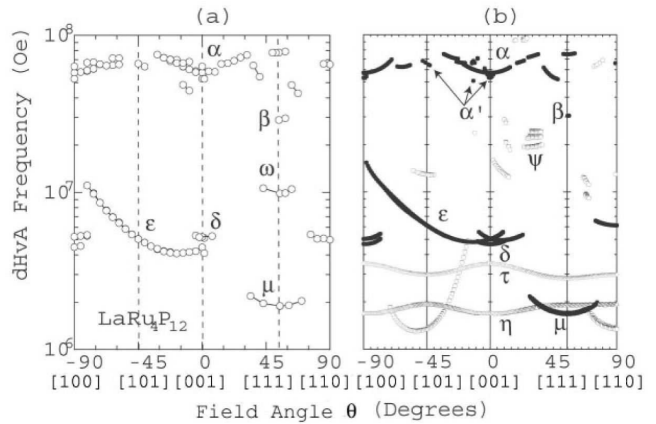}
\caption{Angular dependence of the (a) experimental and (b) theoretical dHvA frequencies in LaRu$_4$P$_{12}$.
The dHvA branches indicated by the closed marks in the Fig.~\ref{dHvA_ang_exp_calc}(b) are identified by the experiments.}
\label{dHvA_ang_exp_calc}
\end{center}
\end{figure}

% \section{Results and discussions}
%
Figure~\ref{Osc_FFT} shows (a) a typical recorder trace of the dHvA oscillations and (b)
the corresponding fast Fourier transformation (FFT) spectra in LaRu$_4$P$_{12}$ at 31 mK
for the magnetic field ($H$) along the
$\langle 111 \rangle$ direction. The dHvA signals have been observed just above the superconducting upper
critical field $H_{\rm C2}\sim$ 36.5 kOe, indicating also the high quality of the sample.
Figure~\ref{dHvA_ang_exp_calc}(a) shows the angular dependences of dHvA frequencies both in the (010)
plane and the (1$\overline 1$0) plane. The calculated dHvA
frequencies are shown in 
Fig.~\ref{dHvA_ang_exp_calc}(b), the description of which is given later.
In the experiment, six fundamental dHvA branches labeled $\alpha$, $\beta$, $\omega$, $\delta$,
$\varepsilon$, and $\mu$ have been observed. The largest frequency branch $\alpha$ is observed
over the whole field angles both in the (010) plane and (1$\overline 1$0) plane,
although it is not perfectly continuous. The $\beta$, $\omega$, and
$\mu$ branches are observed in the limited angular range around [111] direction. The $\delta$
%
%
%%%%%%%%%%%%%%%%%%%
%%%%Figure 1
%%%%%%%%%%%%%%%%%%%
%%%%%%%%%%%%%%%%%%%%%%%%%%%%%%%%%%%%%%%%%%
%%%%%%%%%%%%%%%%%%%%%%%%%%%%%%%%%%%%%%%%%%%%%%%
%
%
branch is observed in a narrow angular range centered at the [001] direction. The $\varepsilon$
branch is observed in a wider angular range in the (010) plane except near [100].

\begin{figure}[hbtp]
%h=here, t=top, b=bottom, p=separate figure page
\begin{center}\leavevmode
\includegraphics[width=7cm]{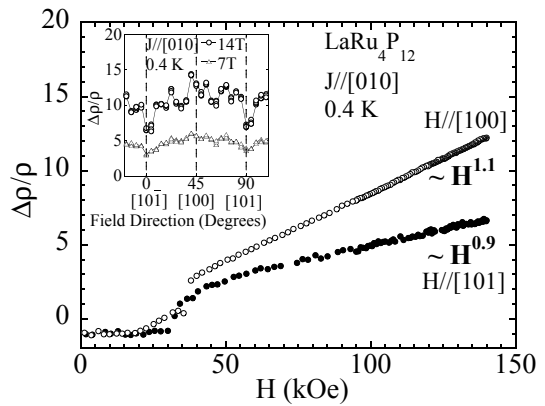}
\caption{Magnetic field dependence of the transverse magnetoresistance (TMR) for $H$ along the [100] and [101] directions.
The inset shows the angular dependence of TMR under the constant fields of 7~T and 14~T in the (010) plane.}
\label{MR}
\end{center}
\end{figure}

We have measured the transverse magnetoresistance (TMR), $\Delta\rho/\rho=[\rho(H)-\rho(0)]/\rho(0)$,
in LaRu$_4$P$_{12}$ to confirm the existence of open orbit on the FS. Figure~\ref{MR} shows
the field dependence of TMR for the field along the [100] and [101] directions and the inset
shows the angular dependences of TMR under the constant fields in the (010) plane. 
%
%
%%%%%%%%%%%%%%%%%%%
%%%%Figure 2
%%%%%%%%%%%%%%%%%%%
%%%%%%%%%%%%%%%%%%%%%%%%%%%%%%%%%%%%%%%%%%%%%%%
%%%%%%%%%%%%%%%%%%%%%%%%%%%%%%%%%%%%%%%%%%%%%%%%%%%%%%
%
%
As was also observed in LaFe$_4$P$_{12}$,~\cite{suga3} the angular dependence of TMR in LaRu$_4$P$_{12}$ is highly anisotropic,
suggesting the existence of an open orbit. Actually,  as shown later, we have observed the multiply connected (48th-band)
hole FS in LaRu$_4$P$_{12}$, which is similar to that of LaFe$_4$P$_{12}$.
LaRu$_4$P$_{12}$ is an uncompensated metal with a different carrier concentration of 
electrons and holes, since it has one molecule per unit cell.
If TMR increases with the field, then an open orbit exists for the direction perpendicular
to both the current and field.
In the present experiment for $H\|[100]$, TMR increases with {\it H}~$^{\rm 1.1}$,
suggesting the existence of an open orbit.
%
%
%%%%%%%%%%%%%%%%%%%
%%%%Figure 3
%%%%%%%%%%%%%%%%%%%
%%%%%%%%%%%%%%%%%%%%%%%%%%%%%%%%%%%%%%%%%%%%%%%
%%%%%%%%%%%%%%%%%%%%%%%%%%%%%%%%%%%%%%%%%%%%%%%%%%%%%%
%
%

\begin{figure}[hbtp]
%h=here, t=top, b=bottom, p=separate figure page
\begin{center}\leavevmode
\includegraphics[width=6.5cm]{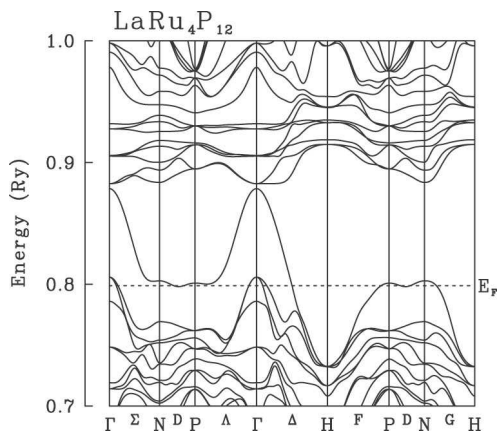}
\caption{Energy band structure of LaRu$_4$P$_{12}$.}
\label{bandW}\end{center}\end{figure}

\begin{figure}[hbtp]
%h=here, t=top, b=bottom, p=separate figure page
\begin{center}\leavevmode
\includegraphics[width=5cm]{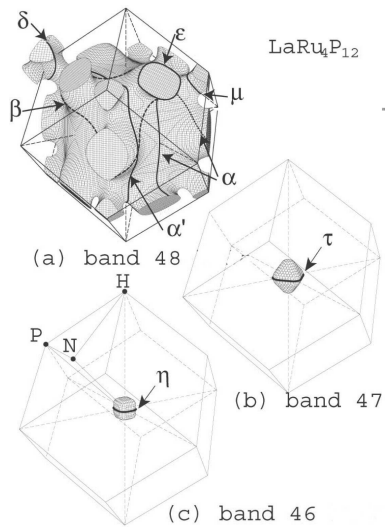}
\caption{The Fermi surface of LaRu$_4$P$_{12}$. The estimated volumes of the
$\tau$- and $\eta$-band FSs are $\sim0.3$\% and $\sim0.2$\% of BZ, respectively.}
\label{FS}\end{center}\end{figure}

In order to assign the origin of dHvA branches, the band-structure calculation has been performed
by using the full potential linearized augmented-plane-wave (FLAPW) method with the local-density
approximation (LDA). We used the room-temperature lattice constant $a=8.0608 {\rm \AA}$ and the fractional
coordinates of P at the 24g site as $x=0$, $y=0.3594$, and $z=0.1434$ for the calculation.~\cite{sekine3}
The details of the calculation are essentially the same as in 
Ref. \onlinecite{suga3}. Figures~\ref{bandW} and \ref{FS} show the
calculated energy band structure and the FS in LaRu$_4$P$_{12}$, respectively.
The FS is composed of the 46th, 47th and 48th band hole sheets. The main difference with
the FS of PrRu$_4$P$_{12}$ is the existence of the 46th and 47th small hole sheets in LaRu$_4$P$_{12}$,
while PrRu$_4$P$_{12}$ has only a multiply connected 49th band hole-FS sheet with one
band counted for two localized 4$f$ electrons.~\cite{harima1} 
The 46th and 47th bands form nearly spherical sheets centered at the $\Gamma$ point,
which stretch slightly  along the $\langle 111 \rangle$ and $\langle 100 \rangle$ directions, respectively.
The 48th band gives a multiply
connected hole-FS sheet centered at the $\Gamma$ point. Its shape is roughly a round cube whose
volume is nearly one half of the BZ size and has a good nesting vector ${\bf q}$= (1,0,0) as expected also in
the 49th-band FS of PrRu$_4$P$_{12}$. 
%
%
%%%%%%%%%%%%%%%%%%%
%%%%Figure 4
%%%%%%%%%%%%%%%%%%%
%%%%%%%%%%%%%%%%%%%%%%%%%%%%%%%%%%%%%%%%%%%%%%%%%%%%%%
%
%

The calculated dHvA frequencies, shown in Fig.~\ref{dHvA_ang_exp_calc}(b), reasonably reproduce the angular dependence of 
the observed dHvA branches $\alpha$, $\beta$, $\delta$, $\varepsilon$, and $\mu$ except $\omega$.
The $\alpha$' branch is found to originate from the electronlike orbits in the 48th band FS.
The other dHvA branches predicted in the band-structure calculation have not been observed in the
present experiments, and the probable origin of this discrepancy is discussed later. 
We have also estimated the cyclotron effective mass $m^{\rm \ast}_{\rm c}$ from the temperature dependence
%
%
%%%%%%%%%%%%%%%%%%%
%%%%Figure 5
%%%%%%%%%%%%%%%%%%%
%%%%%%%%%%%%%%%%%%%%%%%%%%%%%%%%%%%%%%%%%%%%%%%%%%%%%%
%
%
of the dHvA amplitude. The comparison of dHvA frequencies and $m^{\rm \ast}_{\rm c}$ between the
experiment and calculation is given in Table~\ref{table1}.
The effective mass in the experiment is enhanced
roughly twice compared with the calculated one, which is consistent with the enhancement of the
Sommerfeld coefficient ($\gamma$) between the experiment and the theory.~\cite{aoki3} Such a large mass enhancement
in the La compound is also found in LaFe$_4$P$_{12}$.~\cite{suga3} The large mass enhancement in
LaRu$_4$P$_{12}$ should be ascribed to the electron-phonon interaction,
taking into account the rather high value of the superconducting transition temperature
$T_{\rm C}$. Following the McMillan formula
with the value of Coulomb pseudopotential $\mu^*$=0.13,~\cite{mcmillan}
the Debye temperature $\Theta_D =$405 K, and $T_{\rm C} =$7.3 K,~\cite{aoki3}
the electron-phonon coupling constant is estimated to be 0.65
in LaRu$_4$P$_{12}$, which is considered as a moderate-coupling superconductor. 
The value of $\gamma =$44.4 mJ/mol K$^2$ and the jump of specific
heat at $T_{\rm C}$ yield the ratio $\Delta C/\gamma T_{\rm C} \simeq 1.58$, which
is larger than 1.43 of weak-coupling BCS theory.
%
%
%%%%%%%%%%%%%%%%%%%
%%%%Table 1
%%%%%%%%%%%%%%%%%%%
\begin{table*}[hbtp]
\caption{Comparison of the dHvA frequency $F$ and cyclotron effective mass $m_{\rm c}^*$ between
the experiment and the band structure calculation in LaRu$_{4}$P$_{12}$.}
\centering
\begin{tabular}{@{\hspace{\tabcolsep}\extracolsep{\fill}}cccccc}
\hline
\multicolumn{2}{c}{} & \multicolumn{2}{c}{Experiment} & \multicolumn{2}{c}{Theory} \\
Field direction & Branch & $F(\times10^7$~Oe) & $m_{\rm c}^*$/$m_0$ & $F(\times10^7$~Oe) & $m_{\rm c}^*$/$m_0$  \\ 
\hline 
     %$H\| \langle 100 \rangle$ & & & & & &\\
     $H\| [100] $ & $\alpha$  & 5.81  & 2.9 & 5.73  & 1.87  \\		
     & $\delta$ & 0.51  & 3.9 & 0.47  & 2.34\\
\hline 
     %$H\| [110 \rangle$ & & & & & &\\
     $H\| [110] $ & $\alpha$' & 6.40  & - & 6.65  & 4.73  \\		
     & $\epsilon$  & 0.50  & 2.6 & 0.61  & 1.40 \\
\hline 
     %$H\|\langle 111 \rangle$ & & & & & & \\
     $H\| [111]$ & $\alpha$ & 7.70  & 9.0 & 7.50 & 4.73  \\		
     & $\beta$ & 2.89 & 11.8 & 3.05  & 3.8   \\
     & $\omega$ & 0.99 & 3.0 & - & - \\
     & $\mu$ & 0.19 & 3.2 & 0.17 & 1.32 \\    
\hline
\end{tabular}
\label{table1}
\end{table*}
%%%%%%%%%%%%%%%%%%%%%%%%%%%%%%%%%%%%%%%%%%%%%%%%%%%
%
%

The most probable reason why some of the theoretical branches are not observed
in the present experiment, particularly for the large dHvA orbits,
is a strong reduction of the dHvA signals due to their large curvature factors
$A''$ ($A''=|\partial ^2A/\partial k_{\rm H}^2|$, where $A$ is the extremal
cross-sectional area of the FS and $k_{\rm H}$ is the wave-vector component
along the field direction).~\cite{shoenberg}
The rapid change of $A$ as a function of $k_{\rm H}$, i.e., the large value of $A''$,
diminishes the dHvA amplitude for
that extremal area. For the field angle $\theta =$30
degrees: $A''=0.14$ for the $\alpha$ branch whereas $A''=22$ for the $\psi$ branch in the
calculation [see Fig.~\ref{dHvA_ang_exp_calc}(b)].
However, that is not the case for $\tau$ and $\eta$ branches,
which originate from 47$^{th}$- and 46th-band FSs, i.e., for $H\|[111]$,
$A''=0.18$ for the $\tau$ branch and $A''=0.61$ for the $\eta$ branch. 
The experimental limitations, i.e., insufficient temperatures and magnetic fields,
could not be the origin of no observation of the $\tau$ and the $\eta$ branches,
since the $\mu$ branch with almost the same curvature factor ($A''=0.4$)
and the larger effective mass (see table~\ref{table1}) than those of the $\tau$ and $\eta$ branches 
(the calculated effective masses for $H\|[111]$ are 0.34$m_0$ for the $\tau$ branch
and 0.23$m_0$ for the $\eta$ branch) has been clearly observed.
Therefore, dHvA signals should be observed if the $\tau$ and $\eta$ branches really exist.    
The disagreement between the experiment and calculation concerning
the $\tau$ and $\eta$ branches 
may be ascribed to the small difference of the lattice parameter
and/or the effect beyond the LDA treatment.
Note that the band structure calculation shows the absence
of FSs responsible for the $\tau$ and $\eta$ branches in PrRu$_4$P$_{12}$,
though the difference of lattice parameter between LaRu$_4$P$_{12}$ (8.0608 ${\rm \AA}$) and
PrRu$_4$P$_{12}$ (8.0424 ${\rm \AA}$) is quite small ($\sim0.2\%$).~\cite{sekine3}
Thermal expansion measurement suggests that the
lattice constant decreases about 1.2\% at low temperatures ($\leq$20 K) in LaRu$_4$P$_{12}$.~\cite{matsuhira}
The band-structure calculation using the low-temperature lattice parameters 
will be clarified in future work.

From the present dHvA experiment and the band-structure calculation, we have confirmed that
the shape of the main FS of LaRu$_4$P$_{12}$ is like a distorted
cube [see Fig.~\ref{FS}(a)] and its volume is nearly one-half of the first BZ, indicating
the presence of good nesting instability with ${\bf q}=$(1,0,0).
The band calculation in PrRu$_4$P$_{12}$ also
predicts the nearly perfect FS nesting, leading to the M-I transition.~\cite{harima1,harima2}
The obtained topology of the main FS in LaRu$_4$P$_{12}$ is similar to that
by band calculation in PrRu$_4$P$_{12}$. 

Then such a M-I transition should be expected also in LaRu$_4$P$_{12}$,
however none has been reported yet; at least the resistivity and specific heat
measurements show no M-I transition above $T_{\rm C}$.~\cite{uchiumi}
One can infer that the presence of two small spherical $\tau$ and $\eta$ band
FSs predicted by the band calculation, which are absent in PrRu$_4$P$_{12}$,
might suppress M-I transition in LaRu$_4$P$_{12}$.
However, the absence of $\tau$ and $\eta$ branches in the present experiment rules
out such a scenario with large reliability as mentioned above.
In PrRu$_4$P$_{12}$, the change of crystalline structure~\cite{lee1,lee2,hao1}
and the marked increase of resistivity below $T_{\rm MI}$ (Ref. 2) 
are explained by the nesting model, 
suggesting the disappearance of the entire FS  below $T_{\rm MI}$,
based on the band-structure calculations
with the localized 4$f$ electrons.~\cite{harima2,curnoe}
Note that in PrFe$_4$P$_{12}$, which shows an antiferro-quadrupolar transition below 6.5 K,
where only the main part of the FS disappears, the metallic state prevails
and heavy fermion behavior appears.~\cite{sato1,aoki2,suga1}
The inelastic neutron scattering (INS) experiment in PrRu$_4$P$_{12}$ shows sharp
crystal electric field (CEF) excitation peaks below $T_{\rm MI}$ that
gradually broaden across and above $T_{\rm MI}$,~\cite{iwasa} suggesting that the
4$f$ electrons are basically localized, though their hybridization with the
conduction electrons ($c-f$ hybridization) increases to some extent across and above $T_{\rm MI}$.
Cosidering all these facts, naturally, the present experiment suggests an essential
4$f$ electron contribution, which is the main difference between the two compounds,
for the M-I transition.
The $c-f$ hybridization might cooperate with the FS-nesting condition in
causing the M-I transition in PrRu$_4$P$_{12}$,
which explains the absence of the M-I transition in LaRu$_4$P$_{12}$ with no 4$f$ electron contribution.\\  

% \acknowledgments
%
We thank Prof. C.~Sekine, Prof. I.~Shirotani, and Prof. K.~Iwasa for the informative discussion.
One of the authors (S.R.S) acknowledges the support from the Japan Society for the Promotion of Science
(JSPS fellowship for the young scientists). This work was partially supported by a Grant-in-Aid for Scientific Research in Priority
Area "Skutterudite" (No. 15072204 and No. 15072206) and COE Research (10CE2004)
of the Ministry of Education, Culture, Sports, Science and Technology of Japan.

%%%%%%%%%%%%%%%%%%%%%%%%%%%%%%%%%%%%%%%


\begin{thebibliography}{99}
\bibitem{aoki3}Our recent specific-sheat measurement suggests the Sommerfeld coefficient $\gamma \simeq$44.4 mJ/K$^2$mol, while the calculation in the present work gives $\gamma \simeq$19.24 mJ/K$^2$mol.
\end{thebibliography}
\end{document}